\documentclass{iopart}

\usepackage{iopams}
\usepackage{setstack}

\usepackage{graphicx}
\usepackage{color}
\usepackage{times, txfonts}

\definecolor{FGViolet}{rgb}{0.61,0.32,0.61}
\definecolor{FGBlue}{rgb}{0,0,0.8}
\definecolor{FGGreen}{rgb}{0.2,0.7,0.2}
\definecolor{FGRed}{rgb}{0.8,0,0}
\definecolor{FGLightGray}{rgb}{0.85,0.85,0.85}
\definecolor{FGGray}{rgb}{0.6,0.6,0.6}

\newcommand{\linemediumsolid}[1][black]{\unitlength1ex
  ({\color{#1}\begin{picture}(6,1)
  \linethickness{0.4mm}
  \put(0,0.5){\line(1,0){6.0}}
  \end{picture}})\nolinebreak
}

\newcommand{\linemediumdashed}[1][black]{\unitlength1ex
  ({\color{#1}\begin{picture}(6,1)
  \linethickness{0.4mm}
  \put(0,0.5){\line(1,0){1.5}}
  \put(2.2,0.5){\line(1,0){1.5}}
  \put(4.4,0.5){\line(1,0){1.5}}
  \end{picture}})\nolinebreak
}

\newcommand{\symboldiamond}[1][black]{({\color{#1}$\blacklozenge$})}
\newcommand{\symboltriangle}[1][black]{({\color{#1}$\blacktriangle$})}
\newcommand{\symbolbox}[1][black]{({\color{#1}$\blacksquare$})}
\newcommand{\symbolcirclesmall}[1][black]{({\color{#1}\textbullet})}

\newcommand{\eq}[1]{\begin{equation}#1\end{equation}}
\newcommand{\eqmulti}[1]{\begin{eqnarray}#1\end{eqnarray}}

\newcommand{\ket}[1]{\ensuremath{\,|{#1}\rangle}}

\newcommand{\ketbra}[2]{\ensuremath{\,|{#1}\rangle\!\langle{#2}|\,}}
\newcommand{\matrixe}[3]{\ensuremath{\langle{#1}|\,{#2}\,|{#3}\rangle}}

\newcommand{\op}[1]{\ensuremath{\hat{\mathrm{#1}}}}
\newcommand{\adj}[1]{\ensuremath{{{#1}}^{\dag}}}

\newcommand{\aO}{\ensuremath{\op{a}^{\phantom{\dag}}}}
\newcommand{\aaO}{\ensuremath{\op{a}^{\dag}}}
\newcommand{\cO}{\ensuremath{\op{c}}}
\newcommand{\ccO}{\ensuremath{\adj{\op{c}}}}
\newcommand{\nO}{\ensuremath{\op{n}}}

\newcommand{\HO}{\ensuremath{\op{H}}}
\newcommand{\TO}{\ensuremath{\op{T}}}

\newcommand{\rhoO}{\ensuremath{\op{\rho}}}
\newcommand{\rrhoO}{\ensuremath{\adj{\op{\rho}}}}
\newcommand{\idO}{\ensuremath{\op{1}}}

\newcommand{\ii}{\ensuremath{\mathrm{i}}}
\newcommand{\ee}{\ensuremath{\mathrm{e}}}
\newcommand{\dd}{\ensuremath{\mathrm{d}}}

\newcommand{\B}{\ensuremath{\mathrm{B}}}
\newcommand{\F}{\ensuremath{\mathrm{F}}}
\newcommand{\FF}{\ensuremath{\mathrm{FF}}}
\newcommand{\FA}{\ensuremath{\uparrow}}
\newcommand{\FB}{\ensuremath{\downarrow}}

\newcommand{\eqref}[1]{(\ref{#1})}

\begin{document}

\title[Dynamic structure factor of ultracold Bose and Fermi gases]
  {Dynamic structure factor of ultracold Bose and Fermi gases in
  optical lattices}

\author{Robert Roth}

\address{Institut f\"ur Kernphysik, Technische Universit\"at Darmstadt,
  64289 Darmstadt, Germany}

\author{Keith Burnett}

\address{Clarendon Laboratory, University of Oxford,
  Parks Road, Oxford OX1 3PU, United Kingdom}

\date{\today}

\begin{abstract}    
We investigate the dynamic structure factor of atomic Bose and Fermi gases
in one-dimensional optical lattices at zero temperature. The focus is on
the generic behaviour of $S(k,\omega)$ as function of filling and
interaction strength with the aim of identifying possible experimental
signatures for the different quantum phase transitions. We employ the
Hubbard or Bose-Hubbard model and solve the eigenvalue problem of the
Hamiltonian exactly for moderate lattice sizes. This allows us to
determine the dynamic structure factor and other observables directly in
the phase transition regime, where approximation schemes are generally not
applicable. We discuss the characteristic signatures of the various
quantum phases appearing in the dynamic structure factor and illustrate
that the centroid of the strength distribution can be used to estimate the
relevant excitation gaps. Employing sum rules, these quantities can be
evaluated using ground state expectation values only. Important differences
between bosonic and fermionic systems are observed, e.g., regarding the
origin of the excitation gap in the Mott-insulator phase. 
\end{abstract}


\maketitle

\section{Introduction}

Ultracold atomic gases in optical lattices have proven to be an unique
experimental tool to explore the rich physics of quantum phase
transitions. Since the first experimental observation of the superfluid to
Mott-insulator transition in an ultracold Bose gas contained in a
three-dimensional optical lattice \cite{GrMa02a}, the theoretical and
experimental activity in this field has increased dramatically. Recently,
the superfluid to Mott-insulator phase transition was also confirmed in
one-dimensional lattices \cite{StMo04}. From the experimental point of
view these systems offer a remarkable degree of experimental control:
lattice depth and topology, interaction strengths, and filling factors can
be adjusted in a wide range. Furthermore one has the choice between purely
bosonic or  fermionic single- or multi-component systems as well as
systems of mixed statistics. At the same time the diagnostic tools are
outstanding. The matter wave interference pattern after release from the
lattice already provides a wealth of information. The dynamic response of
the system can be probed, e.g., by tilting or modulating the lattice
potential \cite{GrMa02b,StMo04}. Bragg spectroscopy \cite{StIn99,StCh99}
provides another powerful tool to access dynamic properties. 

Also from the theoretical point of view these systems are appealing. They
open a window to correlation dominated phenomena far beyond the mean-field
regime. The most striking ramification of the influence of correlations
are the different quantum phases appearing in Bose and Fermi gases in
lattices. For sufficiently deep lattice potentials the single-band 
Hubbard or Bose-Hubbard model is well suited to study all these phenomena.
Besides the superfluid to Mott-insulator transition for bosonic atoms in
an optical lattice \cite{JaBr98} many other phenomena have been
successfully described within the Bose-Hubbard model. Presently a lot of
attention is devoted to fermionic systems and Bose-Fermi mixtures in
lattices \cite{AlIl03,BuBl03,LeSa04,RoBu04a} with the intention to
investigate  pairing and fermionic superfluidity in lattice systems.

The aim of this paper is to systematically investigate the dynamic
structure factor of bosonic and fermionic lattice systems within the
(Bose-) Hubbard model. The dynamic structure factor, which is directly
accessible in experiment, e.g., through two-photon Bragg spectroscopy
\cite{StIn99,StCh99,BuZo04}, fully characterises the density-density
response. It provides detailed insight into the spectrum of elementary
excitations and into the role correlations for the dynamics of the system.
A significant amount of work has been done for weakly interacting atomic
gases in optical lattices using the Bogoliubov approach
\cite{PiSt03,MeKr03}. In contrast to these studies we are aiming at the
regime of strong interactions and correlations which is not accessible
within mean-field or Bogoliubov-type models \cite{ReBu03}. In order to be
able to describe the system throughout the whole phase diagram, we perform
an exact numerical solution of the eigenvalue problem of the Hubbard
Hamiltonian \cite{RoBu03a, RoBu03c, RoBu04a}. This enables us to track the
change of the dynamic structure factor through the phase transition and
identify signatures of the different quantum phases. 

After setting the formal stage in section \ref{sec:hubbard} we discuss our
results for the dynamic structure factor of the single-component
zero-temperature Bose gas in a one-dimensional optical lattice in section
\ref{sec:lb}. In section \ref{sec:lff} we transfer this to a two-component
Fermi gas and investigate the behaviour of the dynamic density and spin
structure factor in the presence of repulsive and attractive interactions.
Section \ref{sec:summary} summarises our findings.

\section{Hubbard models and the dynamic structure factor}
\label{sec:hubbard}

\subsection{Bose-Hubbard and Fermi-Fermi-Hubbard model}

A single-component Bose gas in a translationally invariant one-dimensional
lattice of sufficient depth is described by the single-band Bose-Hubbard
Hamiltonian \cite{JaBr98,RoBu03c}
\eq{ \label{eq:B_hamiltonian}
  \HO^{\B} 
  = -J \sum_{i=1}^{I} (\aaO_{i+1} \aO_{i} + \mathrm{H.a.})
    + \frac{V}{2} \sum_{i=1}^{I} \nO_i (\nO_i-1) \;,
}
where $\aaO_{i}$ creates a boson in the lowest Wannier state localised at
site $i$ and $\nO_i = \aaO_{i}\aO_{i}$ is the associated occupation number
operator. The first term describes the hopping between adjacent sites with
a tunnelling matrix element $J$. Throughout this paper periodic boundary
conditions are assumed, i.e., the hopping term connects the first and the
last site of the lattice. The second term in \eqref{eq:B_hamiltonian}
introduces an on-site two-body interaction of strength $V$. 

We exactly solve the matrix eigenvalue problem of the Hamiltonian in a
basis of Fock states $\ket{\{n_1,...,n_I\}^{\B}_{\alpha}}$ with $n_i =
0,1,2,...$. The index $\alpha=1,...,D$ labels the different compositions
of occupation numbers $\{n_1,...,n_I\}^{\B}_{\alpha}$ for fixed total particle
number $N=\sum_{i=1}^{I} n_{i}$. The resulting basis dimension is
$D=(N+I-1)!/[N! (I-1)!]$. Through a powerful Lanczos-type algorithm we
compute up to $2500$ energy eigenvalues $E_{\nu}$ at the lower end of the
spectrum. The corresponding  eigenvectors $C_{\alpha}^{(\nu)}$ determine
the eigenstates 
\eq{ \label{eq:expansion}
  \ket{\Psi^{\B}_{\nu}}
  = \sum_{\alpha=1}^{D} C_{\alpha}^{(\nu)} \ket{\{n_1,...,n_I\}^{\B}_{\alpha}}
  \;,
}
which are the starting point for the calculation of all relevant
observables, like mean occupation numbers, occupation number fluctuations,
density matrices, and the dynamic structure factor. 

The analogous model for a two-component system of fermionic atoms in a
one-dimensional lattice is defined by the (Fermi-Fermi-) Hubbard
Hamiltonian
\eq{
  \HO^{\FF} 
  = -J \sum_{i=1}^{I} (\aaO_{\FA i+1} \aO_{\FA i} + \mathrm{H.a.})
    -J \sum_{i=1}^{I} (\aaO_{\FB i+1} \aO_{\FB i} + \mathrm{H.a.})
    + V \sum_{i=1}^{I} \nO_{\FA i} \nO_{\FB i} \;,
}
where $\aaO_{\FA i}$ and $\aaO_{\FB i}$ are creation operators for a
fermionic atom of component $\FA$ and $\FB$, respectively, localised at
site $i$. $\nO_{\FA i}$ and $\nO_{\FB i}$ are the associated occupation
number operators. The first two terms describe the nearest neighbour
hopping with a tunnelling matrix element $J$ assumed to be equal for the two
components. The last term gives the on-site interaction between two atoms
of different components with strength $V$. A on-site interaction between
atoms of the same component is excluded from the outset by 
restricting the model space to a single band.

Again we use a complete basis of Fock states and perform an exact matrix
diagonalisation of the Hamiltonian. For a single fermionic component with
$N$ particles, a basis of Fock states 
$\ket{\{n_1,...,n_I\}^{\F}_{\alpha}}$ is given by all possible
arrangements of $N$ sites with $n_i=1$ and $(I-N)$ sites with $n_i=0$. The
dimension $D=I!/[N! (I-N)!]$ of this basis is much smaller than for a
corresponding bosonic system, since the Pauli principle forbids
multiply occupied sites. The Fock basis of a two-component system is given
by the direct product of all possible combinations of the single-component
basis states $\ket{\{n_1,...,n_I\}^{\F\FA}_{\alpha}} \otimes
\ket{\{n_1,...,n_I\}^{\F\FB}_{\beta}}$. The solution of the matrix eigenvalue
problem of the Hamiltonian in this basis yields eigenstates of
two-component fermion system of the form
\eq{ 
  \ket{\Psi^{\FF}_{\nu}}
  = \sum_{\alpha=1}^{D_{\FA}} \sum_{\beta=1}^{D_{\FB}} 
    C_{\alpha\beta}^{(\nu)} 
    \ket{\{n_1,...,n_I\}^{\F\FA}_{\alpha}}\otimes\ket{\{n_1,...,n_I\}^{\F\FB}_{\beta}}
  \;.
}

\subsection{Dynamic structure factor}

In order to characterise and investigate the dynamic density-density
response of bosonic or fermionic lattice systems to an external probe,
which delivers a momentum $k$ and an energy $\omega$, we employ the
dynamic structure factor $S(k,\omega)$. It is assumed that the coupling
between probe and system is sufficiently weak to render a linear response
treatment applicable. At zero temperature the dynamic structure factor is
defined as
\cite{PiNo66}
\eq{ \label{eq:strucfact_def}
  S(k,\omega)
  = \sum_{\nu} | \matrixe{\Psi_{\nu}}{\rrhoO_k}{\Psi_0} |^2\;
    \delta(\omega - E_{\nu 0}) \;,
}
where $\ket{\Psi_{\nu}}$ are the eigenstates of the many-body Hamiltonian
and $E_{\nu 0} = E_{\nu} - E_0$ are the excitation energies relative to
the ground state. In this section we omit any reference to the bosonic or
fermionic system, since the formal steps are identical for both.

The operator $\rrhoO_k$ describes the way the external probe couples to
the system. In the case of the dynamic structure factor it creates an
excitation with momentum $k$ by promoting particles from a state with
quasimomentum $q$ to a state with quasimomentum $q+k$. In the language of
second quantisation this reads \cite{PiNo66}
\eq{ \label{eq:densityflucop}
  \rrhoO_k
  = \sum_{q=0}^{I-1} \ccO_{q+k} \cO_{q} \;,
}
where the $\ccO_q$ are creation operators associated with quasimomentum
eigenstates, i.e., Bloch states. Within the Hubbard model the $\ccO_q$ can
be rephrased in terms of the creation operators $\aaO_i$ with respect to
the localised Wannier states through
\eq{ \label{eq:quasimomcreation}
  \cO_q
  = \frac{1}{\sqrt{I}} \sum_{i=1}^{I} 
    \ee^{\ii (2\pi/I)\, q\, i}\; \aO_i \;.
}
We employ reduced units for the quasimomenta such that only integer values
of $q$ appear. The first Brillouin zone contains $I$ different
quasimomentum states which are labelled $q=0,...,I-1$ for convenience. The
quasimomenta in standard units are given by $2\pi\,q/(a I)$, where $a$ is
the lattice spacing. Inserting
\eqref{eq:quasimomcreation} into equation \eqref{eq:densityflucop} leads
to the simple expression  
\eq{ \label{eq:fluctuationop}
  \rrhoO_k  
  = \sum_{i=1}^{I} \ee^{-\ii (2\pi/I)\, k\, i}\; \nO_i \;.
}
This form clearly shows that the operator $\rrhoO_k$ is related to a
periodic density modulation with quasimomentum $k$. In order to isolate
the fluctuating part, one might subtract the mean occupation numbers
$\bar{n}_i$ for the ground state from the occupation number operator
$\nO_i$ in \eqref{eq:fluctuationop}. However, this affects only the
trivial $k=0$ behaviour of the dynamic structure factor.

The matrix elements $\matrixe{\Psi_{\nu}}{\rrhoO_k}{\Psi_0}$ entering into
the dynamic structure factor can be calculated explicitly for each excited
state $\ket{\Psi_{\nu}}$ obtained by the diagonalisation of the Hubbard
Hamiltonian. Using the expansion \eqref{eq:expansion} of the eigenstates
in terms of Fock states we obtain
\eqmulti{
  \matrixe{\Psi_{\nu}}{\rrhoO_k}{\Psi_0}
  &= \sum_{i=1}^{I} \ee^{-\ii (2\pi/I)\, k\, i}\;
  \sum_{\alpha,\beta=1}^{D} C_{\alpha}^{(\nu)\star} C_{\beta}^{(0)}
    \matrixe{\{n_1...n_I\}_{\alpha}}{\nO_i}{\{n_1...n_I\}_{\beta}} \\
  &= \sum_{i=1}^{I} \ee^{-\ii (2\pi/I)\, k\, i}\;
  \sum_{\alpha=1}^{D} C_{\alpha}^{(\nu)\star} C_{\alpha}^{(0)}
    n_i^{(\alpha)} \;.
}
Since we only compute the lowest $\nu_{\max}$ eigenstates by the Lanczos
procedure, the excitation energy $E_{\nu_{\max}0}$ of the highest eigenstate 
limits the $\omega$ up to which the structure factor is determined. By
checking the sum rules (cf. section \ref{sec:sumrules}) we make sure that all
relevant contributions are contained in this range.

So far, we have discussed a single component system. In a two-component
system, the probe can couple to the two components in different ways. If
the probe does not distinguish the two components, then the operator of
the total on-site occupation number $\nO_{\FA i} + \nO_{\FB i}$ is the
relevant quantity and enters into the definition of the total density
fluctuation operator \cite{BuZo04}
\eq{ 
  \rrhoO_{\mathrm{D}k}  
  = \sum_{i=1}^{I} \ee^{-\ii (2\pi/I)\, k\, i}\; (\nO_{\FA i} + \nO_{\FB
  i}) \;.
}
With equation \eqref{eq:strucfact_def} this defines the dynamic density 
structure factor $S_{\mathrm{D}}(k,\omega)$. If the probe, on the other 
hand, couples to the total on-site spin, then the difference between 
the two occupation number operators  $\nO_{\FA i} - \nO_{\FB i}$ is
decisive and leads to the spin fluctuation operator
\eq{ 
  \rrhoO_{\mathrm{S}k}  
  = \sum_{i=1}^{I} \ee^{-\ii (2\pi/I)\, k\, i}\; (\nO_{\FA i} - \nO_{\FB
  i}) 
}
and the associated dynamic spin structure factor
$S_{\mathrm{S}}(k,\omega)$. Possible ways of measuring these quantities in
ultracold atomic gases are discussed in ref. \cite{BuZo04}.

\subsection{Sum Rules}
\label{sec:sumrules}

Sum rules provide an outstanding tool in connection with the dynamic
structure factor \cite{PiSt03}. They relate the moments of the dynamic
structure factor to ground state expectation values of different
operators. The simplest sum rule relates the dynamic structure factor
$S(k,\omega)$ to the static structure factor $S(k)$. The integral of
$S(k,\omega)$ over $\omega$, i.e., the zeroth moment of the dynamic
structure factor, is proportional to the static structure factor
\eqmulti{ \label{eq:sumrule_zero}
  \int\dd\omega\;S(k,\omega)
  =\sum_{\nu} | \matrixe{\Psi_{\nu}}{\rrhoO_k}{\Psi_0} |^2 
  \overset{!}{=} \matrixe{\Psi_0}{\rrhoO_k \rhoO_k}{\Psi_0}
  \equiv N S(k) \;.
} 
The first equality follows immediately from the definition of the dynamic
structure factor \eqref{eq:strucfact_def}. The equivalence of the second
and the third expression can be easily shown by making use of the completeness 
of the eigenstates $\ket{\Psi_{\nu}}$ of the Hamiltonian. Inserting 
a unit operator $\idO=\sum_{\nu} \ketbra{\Psi_{\nu}}{\Psi_{\nu}}$ between 
the two fluctuation operators in the third expression immediately leads to
the second expression. In view of this relation it is convenient to define
a normalised dynamic structure factor $\bar{S}(k,\omega)=S(k,\omega)/(N
S(k))$, whose zeroth moment equals $1$ independent of $k$.

The first moment of the dynamic structure factor obeys the
energy-weighted or $f$ sum-rule. For Hubbard-type Hamiltonians with next
neighbour hopping one finds
\eqmulti{ \label{eq:sumrule_one}
  \fl\qquad\int\dd\omega\;\omega\;S(k,\omega)
  = \sum_{\nu} E_{\nu0}\;| \matrixe{\Psi_{\nu}}{\rrhoO_k}{\Psi_0} |^2 
  \overset{!}{=} [\cos(2\pi k/I)-1]\; \matrixe{\Psi_0}{\TO}{\Psi_0}
  \equiv f(k) \;,
} 
where $\TO = -J \sum_{i=1}^{I} (\aaO_{i+1} \aO_i + \aaO_i \aO_{i+1})$ is
the kinetic energy part of the Hamiltonian. Again, the second term follows
directly from the definition of the dynamic structure factor. The
equivalence of the second and third expression can be proven by 
considering the ground state expectation value of the double-commutator
$[[\rhoO_k,\HO], \rrhoO_k]$. Notice, the explicit evaluation of the
double-commutator for the Hubbard Hamiltonian leads to a term
$2\,[\cos(2\pi k/I)-1]\,\TO$ in contrast to the usual expression for
homogeneous Bose or Fermi gases \cite{PiNo66}.

We will make use of these sum rules in two different ways. First, we will
check the numerical results for the dynamic structure factor by computing
the moments of the structure factor and the corresponding ground state
expectation values independently. Thus we make sure that the truncated set
of excited states that was obtained numerically, includes all states which
contribute to the dynamic structure factor, i.e., which have non-vanishing
matrix elements $\matrixe{\Psi_{\nu}}{\rrhoO_k}{\Psi_0}$.  Secondly, we
can utilise sum rules to determine integral properties of the dynamic
structure factor without computing excited states explicitly. We can, for
example, determine the energy of the centroid of the strength distribution,
given by the ratio of the first and the zeroth moment of $S(k,\omega)$,
using ground state expectation values alone. We will come back to this
point at the end of section \ref{sec:lb}.

\section{Single-component Bose gas}
\label{sec:lb}

\begin{figure}
\begin{flushright}
\includegraphics[width=0.80\textwidth]{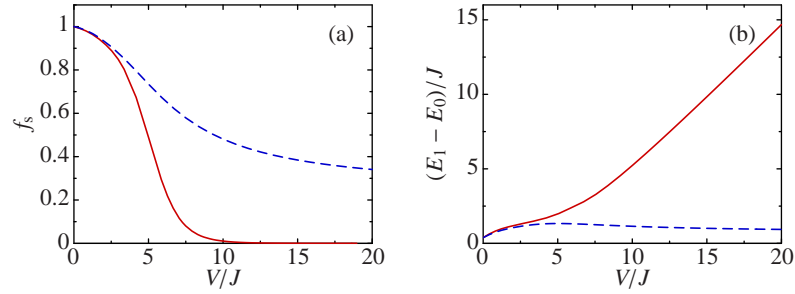}
\end{flushright}
\vspace{-3ex}
\caption{(a) Superfluid fraction $f_{\mathrm{s}}$ as function of the
interaction strength $V/J$ \cite{RoBu03c,RoBu03a} for a commensurate system
with $I=N=10$ \linemediumsolid[FGRed]\ and for non-commensurate filling
$I=10, N=8$ \linemediumdashed[FGBlue]. (b) Energy gap between ground state
and first excited state for the same two cases.}
\label{fig:lb_superfluid}
\end{figure}

The properties of the single-component Bose gas in the presence of
repulsive interactions have been studied in detail experimentally as well
as theoretically. For commensurate filling factors a quantum phase 
transition from a superfluid at low values of $V/J$ to a Mott-insulator 
phase at large $V/J$ appears. For an infinite one-dimensional lattice,
Monte Carlo calculations \cite{BaSc92} and strong -coupling expansion
methods \cite{FrMo96} predict a critical interaction strength
$(V/J)_{\mathrm{crit}}\approx4.65$ for the superfluid to Mott-insulator
transition at $N/I=1$. This is consistent with the recent experimental
observation of the Mott-insulator transition in one-dimensional systems
\cite{StMo04}. We have performed explicit calculations of the superfluid
fraction in finite systems via the phase stiffness \cite{RoBu03c,RoBu03a}.
The behaviour of the superfluid weight $f_{\mathrm{s}}$ and of the energy
gap between ground state and first excited state for a lattice with $I=10$
sites are summarised in figure \ref{fig:lb_superfluid}. For commensurate
filling ($N=10$) the superfluid fraction decreases rapidly and a linearly
growing energy gap emerges. For incommensurate filling ($N=8$) the
superfluid fraction stays finite and the energy gap between ground and
first excited state remains small. A detailed analysis of these properties
as well as other observables was presented in \cite{RoBu03c}.

\begin{figure}
\begin{flushright}
\includegraphics[width=0.87\textwidth]{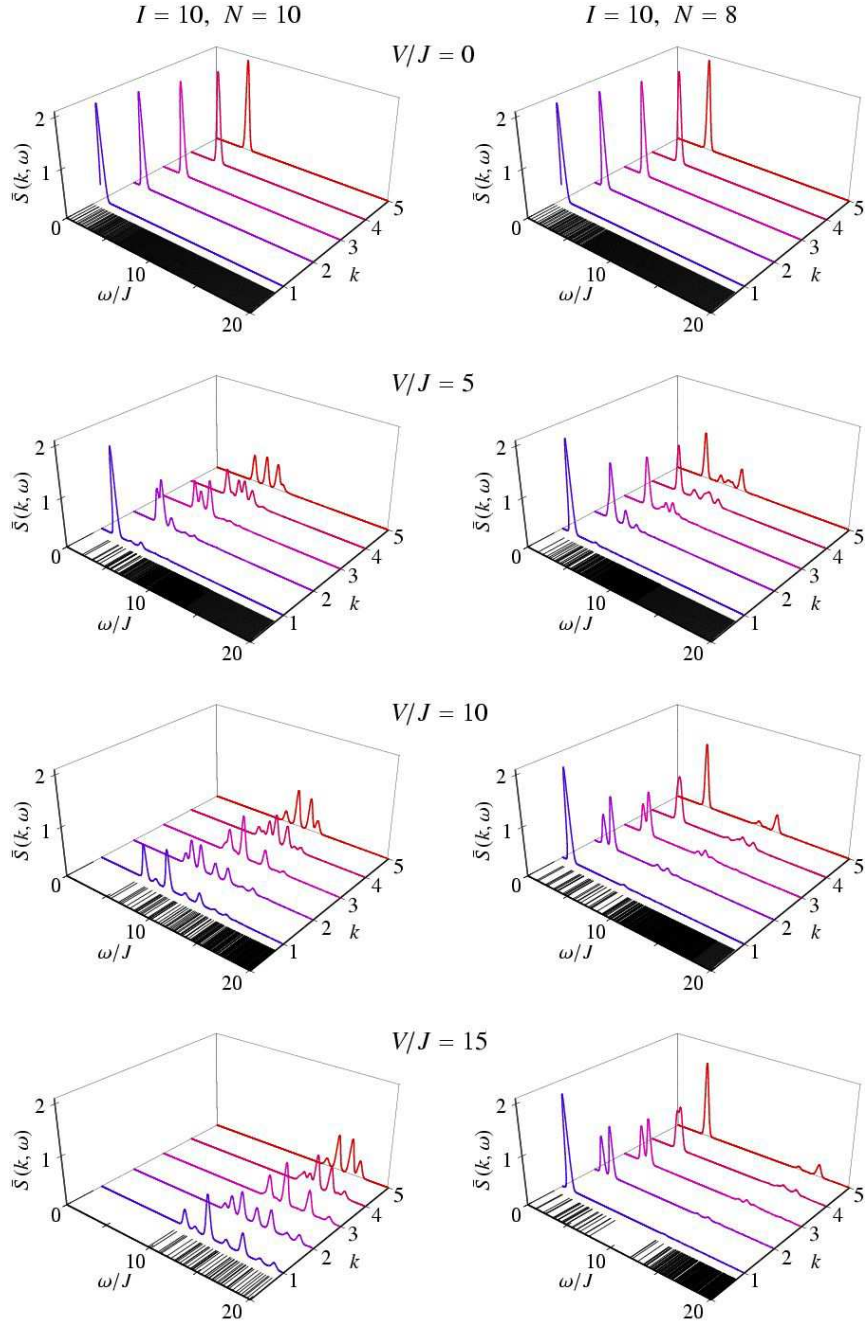} 
\end{flushright}
\caption{Normalised dynamic structure factor $\bar{S}(k,\omega)$ 
for a single-component Bose gas in a one-dimensional lattice with  $I=10$
sites. The plots in the left hand column correspond to a commensurate
filling with $N=10$, the one on the right hand side are for incommensurate
filling with $N=8$. The interaction parameters for the different rows are
$V/J=0$, $5$, $10$, and $15$ (from top to bottom). The bars at the forward
edge of each plot represent the excitation spectrum, i.e. the possible
values of $E_{\nu0}$.}
\label{fig:lb_structfact}
\end{figure}

Our aim is to investigate the change of the dynamic structure factor
$S(k,\omega)$ throughout the whole phase diagram, i.e., for different
interaction strengths $V/J$ and fillings $N/I$. Different characteristic
behaviours might serve as an easily accessible experimental signature to
identify the various quantum phases. 

Figure \ref{fig:lb_structfact} summarises our results for the normalised
dynamic structure factor $\bar{S}(k,\omega)=S(k,\omega)/(N S(k))$ of a
single-component Bose gas with different interaction strengths $V/J$ in a
lattice with $I=10$ sites. For the sake of clarity the $\delta$-function
appearing in the definition \eqref{eq:strucfact_def} of the structure
factor is smoothed out, i.e., it is replaced by a narrow Gauss function
\eq{
  \delta(\omega - E_{\nu 0})
  \to \frac{1}{\sqrt{2\pi}\Delta} 
    \exp\Big[-\frac{(\omega - E_{\nu 0})^2}{2 \Delta^2}\Big]
}
with a width parameter $\Delta/J = 0.2$ (for the fermionic systems
discussed in section \ref{sec:lff} we use $\Delta/J = 0.12$). Each of the
curves shown in the individual plots corresponds to a different value of
the quasimomentum transfer $k$. Only the non-trivial values
$q=1,2,...,I/2$ are shown. For $q=0$ the structure factor vanishes for
$\omega>0$ as can be seen from equation \eqref{eq:strucfact_def}.
Furthermore, due to time-reversal invariance the structure factors for
$-q$ and $q$ are identical. In addition to the dynamic structure factor
each of the plots shows the excitation spectrum as a sequence of levels
along the $\omega$-axis. 

The plots in the left hand column of figure \ref{fig:lb_structfact} show
the dynamic structure factor for commensurate filling $N=I=10$. In the
absence of interactions (upper plot) the excitation strength for each
quasimomentum is concentrated in a single peak at low $\omega$. The peak
shifts slightly towards larger excitation energies when the quasimomentum
transfer $k$ is increased. In the presence of interactions the excitation
strength fragments and spreads over a considerable range of excitation
energies $\omega$ \cite{ReBl04}. This is caused by the change of the
structure of the eigenstates which leads to non-vanishing matrix elements
$\matrixe{\Psi^{\B}_{\nu}}{\rrhoO_k}{\Psi^{\B}_0}$ for several excited
states. 

A more significant change of $S(k,\omega)$ results from the emergence of a
gap in the energy spectrum as soon as one enters the Mott-insulator
phase. Starting from $V/J\gtrsim5$ there is a gap between ground state and
first excited state which grows linearly with $V/J$ (cf. figure
\ref{fig:lb_superfluid}). Hence, there are no low-lying excited states and
therefore no low-lying contributions to the dynamic structure factor within
the Mott-insulator phase. The excitation strength, which is carried by
states just above the energy gap, is pushed towards large $\omega$ as the
gap expands. The plots for $V/J=10$  and $15$ in figure
\ref{fig:lb_structfact} demonstrate this mechanism. 

This can be compared to the behaviour of the dynamic structure factor for
a system with non-commensurate filling ($I=10$, $N=8$) as depicted in the
right hand column of figure \ref{fig:lb_structfact}. As before, the
strength is concentrated in a single low-energy peak for the
noninteracting system. However, with increasing $V/J$ the dominant part of
the strength remains at small $\omega$ since there is no gap between
ground and first excited state. Nevertheless, a small contribution results
from higher lying states which moves to larger $\omega$ with increasing
$V/J$. Interestingly, for strongly repulsive interactions, e.g. $V/J=15$,
the spectrum shows a gap at relatively large excitation energies
$E_{\nu0}/J\approx 10$ and the high-lying strength is carried by states
just above this gap.   

\begin{figure}
\begin{flushright}
\includegraphics[width=0.8\textwidth]{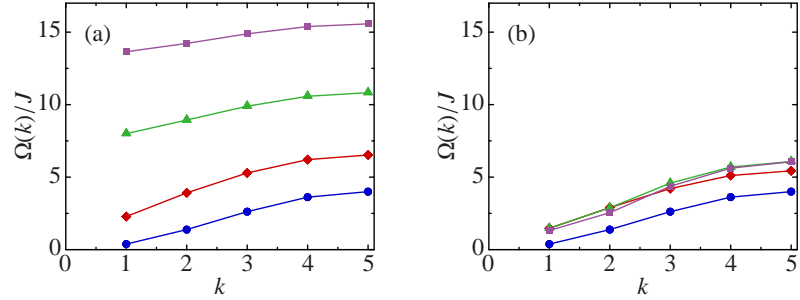}
\end{flushright}
\vspace{-3ex}
\caption{Position of the centroid \eqref{eq:centroid} of the dynamic
structure factor as function of $k$ for a single-component Bose gas with 
(a) $N=10$ particles and (b) $N=8$ particles, resp., in a $I=10$ lattice.
The interaction strengths are $V/J=0$ \symbolcirclesmall[FGBlue], $5$
\symboldiamond[FGRed], $10$ \symboltriangle[FGGreen], and $15$
\symbolbox[FGViolet]. The points are connected by straight lines to guide the
eye.}
\label{fig:lb_centroid}
\end{figure}

The gross structural changes of $S(k,\omega)$ discussed above are well
represented by the position of the centroid of the strength distribution
as function of $k$
\eq{ \label{eq:centroid}
  \Omega(k) 
  = \frac{\int\dd\omega\; \omega\; S(k,\omega)}{\int \dd\omega\; S(k,\omega)}
  = \int\dd\omega\; \omega\; \bar{S}(k,\omega)
  = \frac{f(k)}{N S(k)} \;,
}
which is the ratio of the energy-weighted sum rule $f(k)$ defined in
equation \eqref{eq:sumrule_one} and the static structure factor $N S(k)$
given by \eqref{eq:sumrule_zero}. By using the sum-rule
identities given in section \ref{sec:sumrules}, the position of the
centroid can be expressed in terms of ground state expectation values
only. Hence the sum rules enable us to draw conclusions about the
structure of the excitation spectrum once we know the ground state.
Obviously this is of great practical importance, we do not have to compute
the full excitation spectrum if we are interested in gross properties like
$\Omega(k)$.

Figure \ref{fig:lb_centroid} depicts the position of the centroid as
function of $k$ for different interaction parameters $V/J$ and
commensurate as well as non-commensurate filling. In general $\Omega(k)$
decreases with decreasing $k$, however, its minimum value and its
behaviour in the limit of small $k$ is crucial. As long as the system is
in the superfluid phase, $\Omega(k)$ tends towards zero for in the limit
of small $k$. Only within the Mott-insulator phase, i.e., for
$V/J\gtrsim5$ and commensurate filling, the position of the centroid tends
to a finite value for $k\to0$. The minimum value of $\Omega(k)$ can be
used as an estimate for the excitation gap, which, for the bosonic system,
coincides with the gap in the excitation spectrum.

\section{Two-component Fermi gas}
\label{sec:lff}

The qualitative behaviour of the dynamic structure factor for a Bose gas
in the lattice can be largely understood from the structure of the
excitation spectrum alone. This is not the case for two-component Fermi
gases, where the transition matrix elements
$\matrixe{\Psi^{\FF}_{\nu}}{\rrhoO_k}{\Psi^{\FF}_0}$ play a decisive role.

\subsection{Repulsive interactions}
\label{sec:lff_rep}

We first consider the case of repulsive inter-component interactions. For
$N_{\FA}=N_{\FB}=I/2$, i.e., half-filling of the band, there exists a
(Hubbard-) Mott-insulator phase similar to the bosonic system. The
one-dimensional Hubbard model at exactly half-filling is special in  many
respects. It is well known \cite{Giam04} that there exists a mapping
between the charge or density and the spin sector of the model. By
performing a gauge transformation $\aaO_{\FA i} \to (-1)^{i}\,\aO_{\FA i}$ on
one of the components the density observables (e.g. the density structure factor)
of the Hubbard model with repulsive interactions $V>0$ are mapped
onto the corresponding spin observables (e.g. the spin structure
factor) of a system with attractive interaction $-V$ and vice versa.
However, this symmetry does not hold away from half-filling.

\begin{figure}
\begin{flushright}
\includegraphics[width=0.87\textwidth]{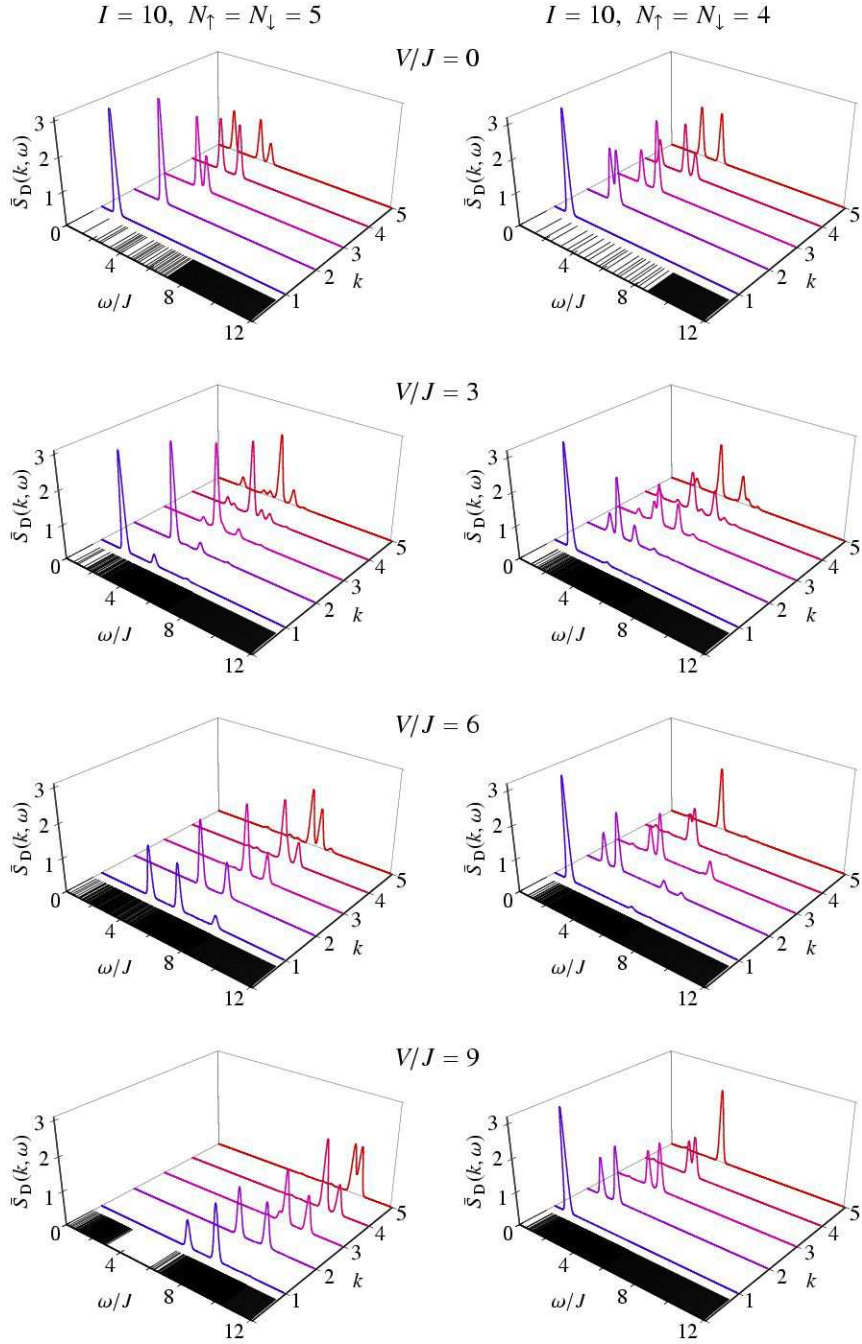} 
\end{flushright}
\caption{Normalised dynamic density structure factor $\bar{S}_{\mathrm{D}}(k,\omega)$ 
for a two-component Fermi gas with repulsive interactions in a $I=10$ lattice. 
The plots in the left hand column show the half-filling case $N_{\FA}=N_{\FB}=5$, 
the ones in the right hand column are for $N_{\FA}=N_{\FB}=4$. The
interaction parameters for the different rows are $V/J=0$, $3$, $6$,
and $9$ (from top to bottom). The bars at the forward edge of each plot 
represent the excitation spectrum, i.e. the possible values of $E_{\nu0}$.}
\label{fig:lff_structfact_rep}
\end{figure}

In the following we investigate the two-component Fermi gas with repulsive
interactions at half-filling of the band ($I=10$, $N_{\FA}=N_{\FB}=5$) and
for particle numbers away from half-filling ($I=10$, $N_{\FA}=N_{\FB}=4$)
and compare with the bosonic systems of section \ref{sec:lb}. The energy
spectra and the normalised dynamic density structure factors
$\bar{S}_{\mathrm{D}}(k,\omega) = S_{\mathrm{D}}(k,\omega)/
[(N_{\FA}+N_{\FB}) S_{\mathrm{D}}(k)]$ for the two-component Fermi gas are
summarised in  figure \ref{fig:lff_structfact_rep}. 

The energy spectrum at half-filling develops a gap with increasing strength
of the repulsive interaction (cf. left hand column of figure
\ref{fig:lff_structfact_rep}). In contrast to the bosonic system at
commensurate filling, the gap does not emerge between ground and first
excited state but at relatively large excitation energies of $E_{\nu0}/J
\approx 5$. Therefore a significant number of states remains below the gap,
and it is a priori not clear whether these contribute to the dynamic
structure factor or not. The explicit calculation of
$S_{\mathrm{D}}(k,\omega)$ reveals a behaviour similar to the bosonic
system at commensurate filling: The strength is shifted towards larger
$\omega$ with increasing interaction strength $V/J$. As the gap appears,
only those states above the gap contribute to the structure factor. All
states below the gap possess vanishing transition matrix elements 
$\matrixe{\Psi^{\FF}_{\nu}}{\rrhoO_{\mathrm{D}k}}{\Psi^{\FF}_0}$ and,
therefore, do not contribute. Hence, the mechanism which generates the
density excitation gap is more complicated than for the bosonic system,
where it was a simple consequence of the gap in the eigenspectrum. 

For systems away from half-filling (right hand column in figure
\ref{fig:lff_structfact_rep}), no corresponding gap develops in the energy
spectrum. Similar to the non-commensurate bosonic system, the dominant
strengths remains at low excitation energies independent of the
interaction strength $V/J$ and only a few marginal peaks are shifted to
higher $\omega$.

\begin{figure}
\begin{flushright}
\includegraphics[width=0.8\textwidth]{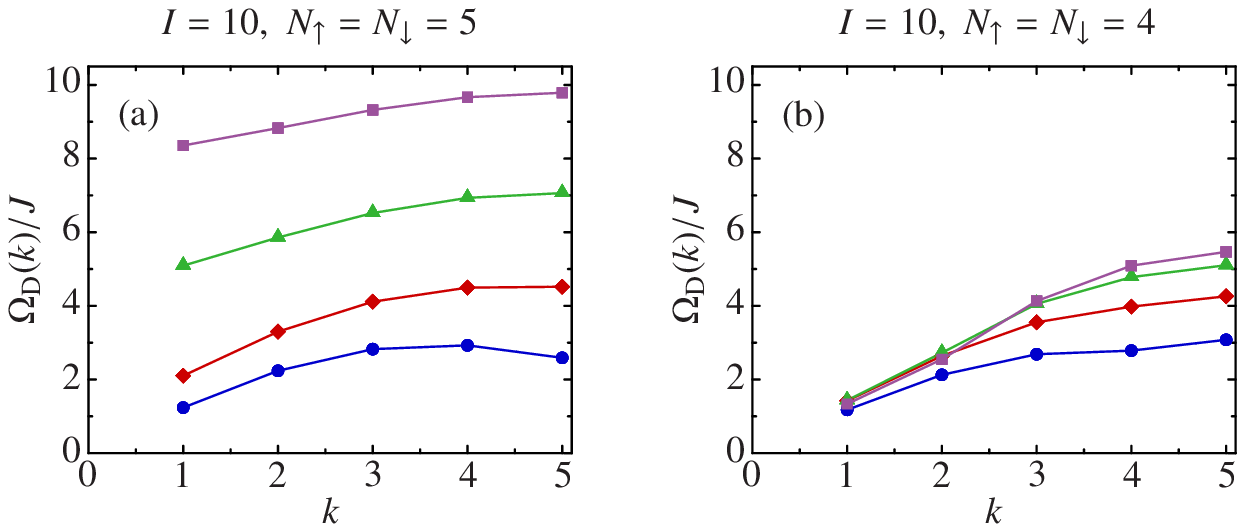}
\end{flushright}
\vspace{-3ex}
\caption{Position of the centroid $\Omega_{\mathrm{D}}(k)$ of the dynamic
density structure factor as function of $k$ for a two-component Fermi gas with 
$N_{\FA}=N_{\FB}=5$ (a) and $N_{\FA}=N_{\FB}=4$ (b), 
resp., in a $I=10$ lattice. Results are shown for repulsive interactions 
with $V/J=0$ \symbolcirclesmall[FGBlue], $3$ \symboldiamond[FGRed], 
$6$ \symboltriangle[FGGreen], and $9$ \symbolbox[FGViolet]. 
The points are connected by straight lines to guide the
eye.}
\label{fig:lff_centroid_rep}
\end{figure}

As for the bosonic system, these observations can be quantified in terms
of  centroid $\Omega_{\mathrm{D}}(k)$ of the dynamic density structure
factor  $S_{\mathrm{D}}(k,\omega)$. The plots in figure
\ref{fig:lff_centroid_rep} display the position of the centroid as
function of $k$ for repulsive interactions of different strengths. At
half-filling the centroid shifts towards larger energy transfers with
increasing $V/J$. As for the bosonic system, the $k$-dependence of
$\Omega_{\mathrm{D}}(k)$ can serve as an indicator for the insulating
state. If $k$ is decreased towards zero, then the centroid tends to zero
in the conducting phase. In the Mott-insulator phase, however, it remains
at a finite value which grows linearly with increasing $V/J$. Again we
point out that this behaviour is non-trivial in the fermionic case, it is
not simply a consequence of the structure of the energy spectrum. Many
states exist below the gap, however, they do not contribute to the dynamic
density structure factor. This entails that the excitation gap, i.e., the
minimum energy transfer $\omega$ required to excite a density fluctuation,
is significantly larger than the gap in the spectrum of the Hamiltonian.  

Away from half-filling (right panel in figure \ref{fig:lff_centroid_rep})
the position of the centroid always tends towards
$\Omega_{\mathrm{D}}(k)/J\approx 0$ for $k\to0$. This is consistent with
our findings for the single-component Bose gas. Hence, the Mott-insulator
leaves a clear and general signature in the dynamic density structure
factor, which is directly accessible to experiment and allows a
distinction from conducting or superfluid phases.

\subsection{Attractive interactions}

Finally, we address the case of attractive interactions in the
two-component Fermi system. This case is of particular interest in view
of pairing and superconductivity in these lattice systems. 

\begin{figure}
\begin{flushright}
\includegraphics[width=0.87\textwidth]{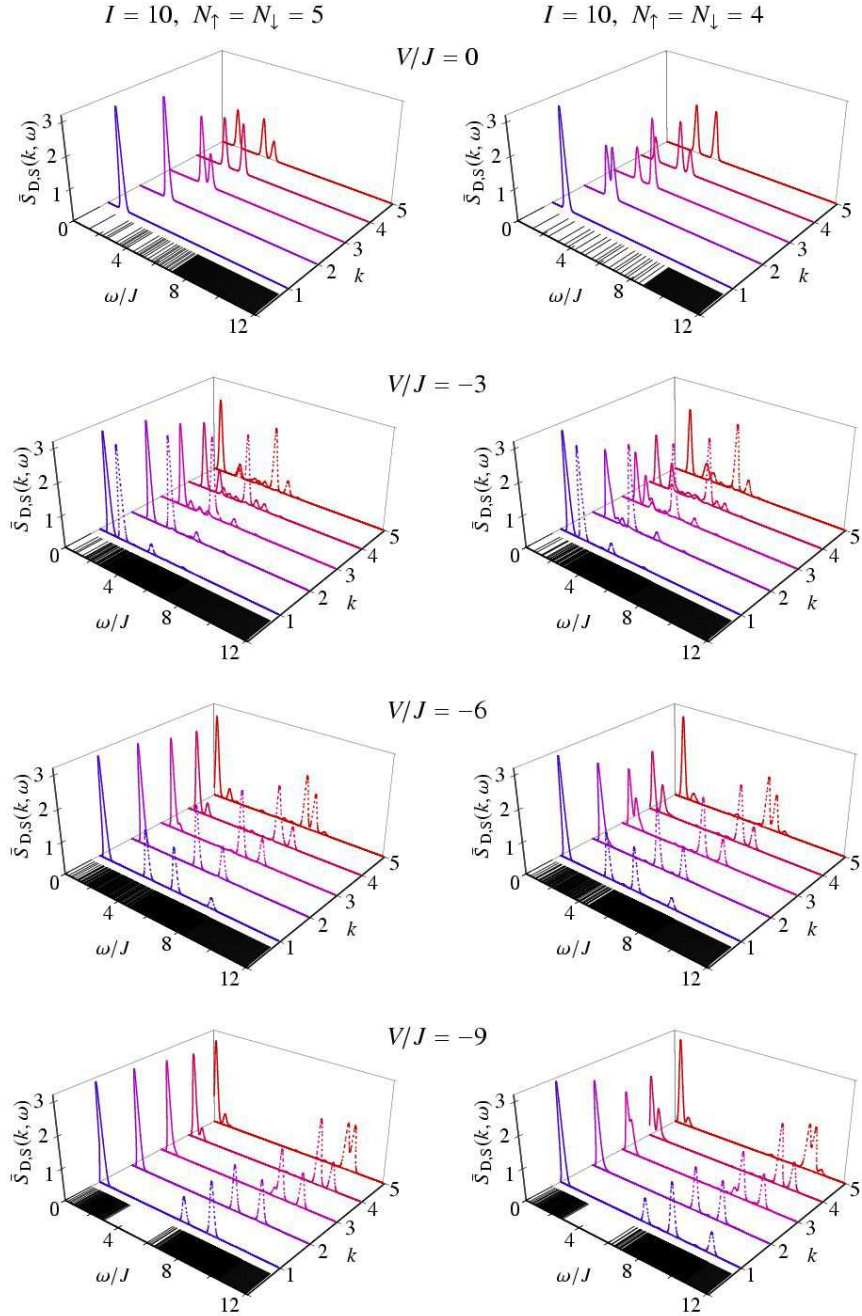} 
\end{flushright}
\caption{Normalised dynamic density structure factor $\bar{S}_{\mathrm{D}}(k,\omega)$
(solid lines) and dynamic spin structure factor $\bar{S}_{\mathrm{S}}(k,\omega)$
(dotted lines) for a two-component Fermi gas with attractive 
interactions in a $I=10$ lattice. 
The plots in the left hand column show the half-filling case $N_{\FA}=N_{\FB}=5$, 
the ones in the right hand column are for $N_{\FA}=N_{\FB}=4$. The
interaction parameters for the different rows are $V/J=0$, $-3$, $-6$,
and $-9$ (from top to bottom). The bars at the forward edge of each plot 
represent the excitation spectrum.}
\label{fig:lff_structfact_attr}
\end{figure}
\begin{figure}
\begin{flushright}
\includegraphics[width=0.8\textwidth]{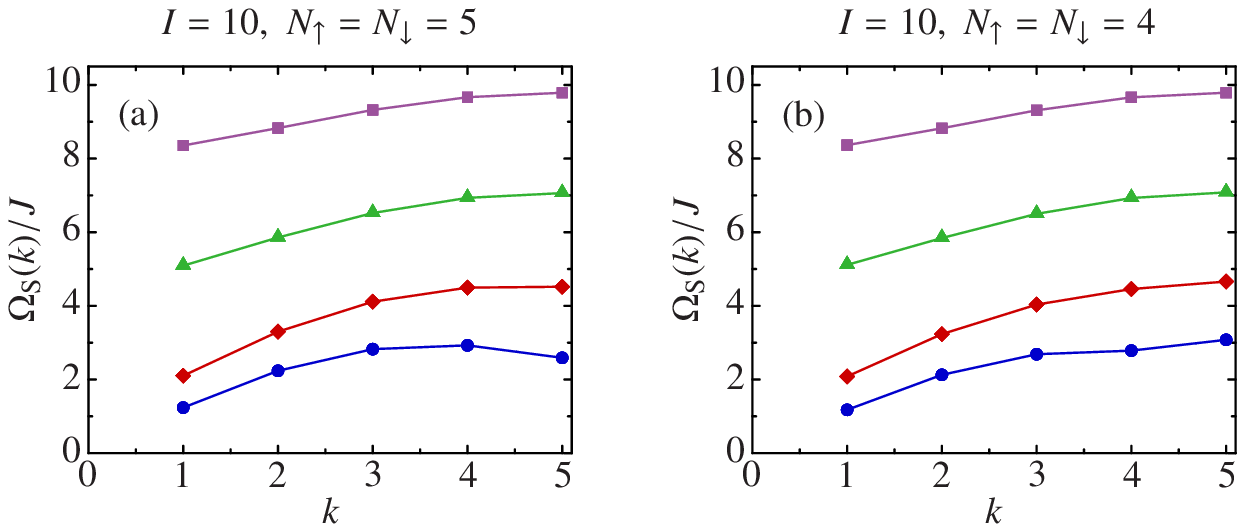}
\end{flushright}
\vspace{-3ex}
\caption{Centroid $\Omega_{\mathrm{S}}(k)$ of the dynamic
spin structure factor for a two-component Fermi gas with 
$N_{\FA}=N_{\FB}=5$ (a) and $N_{\FA}=N_{\FB}=4$ (b), 
resp., and attractive interactions $V/J=0$ \symbolcirclesmall[FGBlue], 
$-3$ \symboldiamond[FGRed], $-6$ \symboltriangle[FGGreen], and 
$-9$ \symbolbox[FGViolet].}
\label{fig:lff_centroid_attr}
\end{figure}

For half-filling the properties of the gas with attractive interactions
can be deduced from the repulsive case on the basis of the spin-density mapping
discussed in section \ref{sec:lff_rep}. Its simplest ramification is
that the excitation spectrum of a two-component Fermi system with
attractive interactions $V<0$ is \emph{identical} to a system with a
repulsive interaction of strength $|V|$. The dynamic structure factor is
also governed by this symmetry: The dynamic \emph{density} structure
factor $S_{\mathrm{D}}(k,\omega)$ of a system with attractive interaction is
identical to the dynamic \emph{spin} structure factor $S_{\mathrm{S}}(k,\omega)$ for
a gas with a repulsive interaction of same magnitude. 

The numerical results for the normalised dynamic density structure factor
$\bar{S}_{\mathrm{D}}(k,\omega)$ of the two-component fermionic system with
attractive interactions are displayed in figure
\ref{fig:lff_structfact_attr} (solid lines). The behaviour of the
excitation strength is in stark contrast to the case of repulsive
interactions. With growing attraction the strength is increasingly
concentrated in one or two peaks at very low excitation energies.  This
effect does not depend on the filling as the comparison of the
half-filling case (left hand column of figure
\ref{fig:lff_structfact_attr}) with a system away from half filling (right
hand column) illustrates. The energy spectra are also largely independent
of filling, for both fillings an energy gap appears for sufficiently strong
attractive interactions. Remember, away from half-filling no mapping
between density and spin sector exists anymore. The spectrum exhibits a
gap for strong attractive interactions but no gap for repulsive
interactions (cf. lower right hand panels in figures
\ref{fig:lff_structfact_rep} and \ref{fig:lff_structfact_attr}). 

The appearance and the physical interpretation of the energy gap is
related to pairing of the fermionic atoms due to the attractive
interaction. This relation is revealed by the dynamic spin structure
factor $S_{\mathrm{S}}(k,\omega)$. As discussed in Ref. \cite{BuZo04} for a
homogeneous two-component Fermi gas, contributions to the spin structure 
factor arise from processes which break the pairs. Hence, the spin
structure factor provides direct information on the pairing gap in the
superconducting state. The dotted lines in figure
\ref{fig:lff_structfact_attr} depict the spin structure factor. It is
clearly seen that with increasing attraction the contributions to 
$S_{\mathrm{S}}(k,\omega)$ are shifted towards larger energy transfers and that no
low-lying strength remains. Eventually, as a gap in the energy spectrum
appears, only those states above the gap contribute to the spin structure
factor. As for the fermionic Mott-insulator the absence of low-lying
strength implies the vanishing of the transition matrix elements 
$\matrixe{\Psi^{\FF}_{\nu}}{\rrhoO_{\mathrm{S}k}}{\Psi^{\FF}_0}$ for all
states below the gap. Notice, the spin-excitation gap is significantly larger than
gap in the energy spectrum and it already appears for interaction strengths
where the spectrum does not yet exhibit a gap.

We can use the centroid $\Omega_{\mathrm{S}}(k)$ of the dynamic spin
structure factor to characterise the gross features of spin excitations on
the basis of sum rules. As for the Mott-insulator phase we can assess the
existence and size of a spin-excitation gap already from the
$k$-dependence of $\Omega_{\mathrm{S}}(k)$ without ever calculating
excited states. Figure \ref{fig:lff_centroid_attr} depicts the
$k$-dependence of $\Omega_{\mathrm{S}}(k)$ for the two fillings considered
previously. Independent of the filling the centroid of the spin structure
factor clearly shows the appearance of the spin-excitation gap already for
weakly attractive interactions. As for the Mott-insulator we can use its
behaviour for $k\to0$ to estimate the size of the excitation gap.

At half-filling the centroid of the spin structure factor depicted
in panel (a) is identical to the centroid of the density structure factor
for repulsive interactions of same magnitude (cf. figure
\ref{fig:lff_centroid_rep}) by virtue of the spin-density mapping.
Away from half-filling this mapping does not apply and the behaviour of
the centroids is completely different.

\section{Summary}
\label{sec:summary}

We have used the dynamic structure factor, a direct experimental
observables, to pin-down signatures of the different zero-temperature 
quantum phases of single-component Bose and two-component Fermi gases in
optical lattices. The systems are described within the single-band (Bose-)
Hubbard model and we perform an exact diagonalisation of the Hamiltonian
using Lanczos methods. Since we are interested mainly in generic
signatures the computational restriction of this approach to moderate
lattice sizes is of minor relevance. The important benefit of this exact
approach is that all regions of the phase diagram can be addressed on the
same footing. 

The comparison of the full dynamic structure factor $S(k,\omega)$ for
different interaction strengths and fillings exhibits clear signatures of
the different quantum phases. The Mott-insulator phases in the bosonic and
fermionic systems are associated with a finite excitation gap which grows
linearly with $V/J$. That is, excitations in form of density fluctuations
are possible only for energies $\omega$ larger than the excitation gap.
Whereas the excitation gap in the bosonic system is directly determined by
the gap between ground state and first excited state of the Hamiltonian,
its origin in the two-component Fermi system is more involved. Many
eigenstates of the Hamiltonian are available in the low-$\omega$ regime,
however, they do not contribute to the excitation strength since their
transition matrix elements for the fluctuation operator vanish. Hence, a
clear distinction between excitation gap (for a particular type of
excitation) and the gap in the eigenspectrum of the Hamiltonian is
indispensable.

For the two-component Fermi gas with attractive interactions the dynamic
spin structure factor indicates the onset of pairing between atoms of the
different components. Similar to the density structure factor for the
Mott-insulator phase, the spin structure factor exhibits an excitation
gap. No spin fluctuations, which would require the breaking of a pair, are
possible below an energy $\omega$ corresponding to the spin excitation
gap. This phenomenon is independent of the total filling --- unlike the
Mott insulator --- but requires equal population of the two components. 
The detailed nature of the pairing in the presence of attractive
interactions will be discussed elsewhere.

Most of the physics revealed by the dynamic structure factors is already
contained in integral quantities like the centroid of the strength
distribution, which can be evaluated using sum rules. The knowledge of the
ground state alone is sufficient to assess the appearance and size of
excitation gaps. The explicit calculation of the excitation spectrum is
required only if details of the dynamic structure factors are of
interest.  This paves the way towards predictions for larger system, e.g.,
within a Monte Carlo or Density Matrix Renormalisation Group approach,
where the computation of the full excitation spectrum is not feasible.

\section*{Acknowledgements}

This work was supported by the DFG, the UK EPSRC, and the EU via the
``Cold Quantum Gases'' network.  K.B. thanks the Royal Society and Wolfson
Foundation for support. R.R. acknowledges fruitful discussions with Karen
Braun-Munzinger, Jacob Dunningham, Markus Hild, and Felix Schmitt.



\end{document}